\documentclass[preprint,12pt]{elsarticle}

\usepackage{amssymb}
\usepackage{amsmath}
\usepackage{mathtools}
\usepackage{listings}
\usepackage{subcaption}
\usepackage[draft]{minted}

\usepackage{cprotect}

\usepackage[hyphens]{url}
\usepackage{hyperref}
\hypersetup{breaklinks=true}

\newcounter{bla}

\usepackage{algorithm}
\usepackage{algpseudocode}
\journal{Computer Physics Communications}

\begin{document}

\begin{frontmatter}



\title{OpenBTE: a Solver for\textit{
ab-initio} Phonon Transport in Multidimensional Structures}


\author[a]{Giuseppe Romano}
\cortext[author] {Corresponding authors.\\\textit{E-mail address:}romanog@mit.edu}
\address[a]{Massachusetts Institute of Technology, 77 Massachusetts Avenue, Cambridge, Massachusetts 02139, USA}

\begin{abstract}

Controlling heat flow at the nanoscales is pivotal to several applications, including thermal energy harvesting and heat management. However, engineering nanostructures is challenging because phonon-boundary interaction, not contemplated by Fourier's law, must be taken into account. Nondiffusive models, such as the Boltzmann transport equation (BTE), have been successfully employed to capture size effects in complex structures; however, their widespread has been hindered by the limited offer of open-source solvers in this space. We fill this void by introducing \verb!OpenBTE!, an efficient solver for the steady-state phonon BTE in multidimensional structures. This tool is interfaced to \textit{first-principles} calculations, thus it unlocks the calculations of thermal-related properties with no fitting-parameters. As an example, we employ \verb!OpenBTE! to compute the temperature and flux maps, as well as the mode-resolved, effective thermal conductivity of Si membranes with infinite and finite thickness. By unlocking fast nanoscale heat transport simulations, \verb!OpenBTE! may help accelerate the design of nanomaterials for thermal energy applications.

\end{abstract}

\begin{keyword}
Thermal transport, nanostructures

\end{keyword}

\end{frontmatter}

{\bf PROGRAM SUMMARY}

\begin{small}
\noindent
{\em Program Title: \verb!OpenBTE!}                                          \\
{\em Developer's repository link:} https://github.com/romanodev/openbte.git \\
{\em Licensing provisions:} MIT  \\
{\em Programming language: \verb!Python!}                                   \\
{\em Nature of problem:} \textit{Ab-initio} calculation of thermal transport properties, including temperature and heat flux maps, as well as the effective thermal conductivity of multidimensional nanostructures. \\
{\em Solution method:} We implement the space-dependent, \textit{ab-initio} Boltzmann transport equation (BTE) for phonons. The \textit{first-principles} calculation of phonon group velocities, frequencies and scattering times, is delegated to external packages to which \verb!OpenBTE! is interfaced. The current implementation focuses on a flavor of the BTE that interpolates phonon populations onto their vectorial mean-free-paths, allowing for fast and accurate thermal transport simulations.\\
{\em Additional comments:} Vectorization via \verb!NUMPY! and parallelization (including with shared memory) with \verb!MPI4Py!.
 
\end{small}

\section{Introduction}
\label{intro}

The increasing accuracy of \textit{first-principles} calculations of heat transport, along with the growing availability of computing resources, has enabled the systematic prediction of lattice thermal conductivity in semiconductors and two-dimensional (2D) materials~\cite{lindsay2019perspective,lindsay2018survey,mingo2014ab}. If key theoretical developments have solidified our understanding of phonon dynamics at the nanoscales, the fast progress of this field has been unarguably aided by the release of open-source packages. For example, \verb!ShengBTE!~\cite{li2014shengbte} implements the iterative solution of the Boltzmann transport equation (BTE)~\cite{omini1995iterative}, with force constants computed by \textit{first-principles}~\cite{broido2007intrinsic,esfarjani2011heat}. The tool can be primarily used for either bulk materials or nanowires~\cite{li2012thermal}. Later iterations include ~\verb!AlmaBTE!~\cite{Carrete2017AlmaBTEMaterials}, which, among several improvements, features a Monte Carlo solver for superlattices, and \verb!FourPhonons!~\cite{han2021fourphonon,feng2016quantum}, a tool that, as the name suggests, can handle four-phonon scattering. Another notable package is~\verb!Phono3Py!~\cite{phono3py}, where the BTE is solved both within the relaxation time approximation (RTA) and including the full scattering operator~\cite{Chaput2013DirectEquation}. Albeit not strictly required, these tools are mostly used in tandem with super-cell approaches. When a density functional perturbation theory (DFPT) is available for a given material, it is often convenient to compute directly the Fourier-transformed force constants, without resorting to supercells; a prominent package in this space is \verb!d3q!~\cite{paulatto2013anharmonic}, implemented on top of \verb!QUANTUM ESPRESSO!~\cite{giannozzi2009quantum}. This software also includes the possibility of simulating thin films~\cite{paulatto2020thermal} and coherence effects~\cite{simoncelli2019unified}. In a similar effort, the recently released~\verb!Kaldo!\cite{barbalinardo2020efficient} combines the Green-Kubo model and the BTE to take into account material disorder~\cite{isaeva2019modeling}. 

While the dissemination of these software manifests the increasing excitement around phonon related applications, their focus is primarily on bulk materials or simple geometries. In fact, most phonon dynamics simulations in complex geometries are currently performed by MonteCarlo solvers~\cite{landon2014deviational,peraud2011efficient}, with a recent implementation provided by \verb!MCBTE!~\cite{pathak2021mcbte}. Thus, an open-source tool that solves space-dependent heat transport deterministically, i.e. in the same spirit as in bulk calculations, is still pending. We fill this gap by introducing \verb!OpenBTE!, a solver for the \textit{first-principles}, multidimensional phonon BTE. Key capabilities include the calculation of the temperature and heat flux maps, as well as the mode-resolve effective thermal conductivity. \textit{Ab-initio} harmonic and anharmonic properties are delegated to external software. While flexible and straightforward to expand, \verb!OpenBTE! currently implements the anisotropic-mean-free-path BTE (aMFP-BTE)~\cite{romano2021}, a method based on the interpolation of the phonon populations onto the vectorial MFP space. This method has shown a 50x speed up with respect to the mode-resolve BTE for thick Si membranes, while not compromising on the accuracy. Furthermore, it scales with constant time with respect to the number of phonon branches, therefore enabling fast simulations of size effects in complex-unit-cell materials, such as bismuth telluride.

The paper is organized as follows. First, we review the model and the assumptions therein, then we provide details on the implemented algorithm and the software's main modules. An example of a porous Si membrane will follow, and final remarks conclude the paper. We also provides technical implementation details in the appendices. Enabling fast simulations of heat transport in nanostructures, \verb!OpenBTE! sets out to accelerate the development of thermal-related applications, such as thermal management and thermoelectrics. \verb!OpenBTE! is released under the \verb!GPL-3.0! license and currently hosted on GitHub~\cite{openbte}. 

\section{Model}
\label{model}

The steady-state, linearized phonon BTE in the temperature formulation reads~\cite{romano2021}
\begin{equation}\label{bte}
  - C_\mu \mathbf{v}_\mu \cdot\nabla \Delta T_\mu(\mathbf{r}) = \sum_\nu W_{\mu\nu} \Delta T_\nu(\mathbf{r}),
\end{equation}
where $\mu,\nu$ collectively indicate phonon branch and wave vector, running up to $N_p$ and $N_q$, respectively. The unknowns $\Delta T_\mu (\mathbf{r})$ are the phonon pseudo-temperatures (referred to hereafter simply as temperatures), connected to the deviational non-equilibrium phonon population via $\Delta T_\mu(\mathbf{r}) = \Delta n_\mu(\mathbf{r}) \hbar \omega_\mu C_\mu^{-1}$.  The mode-specific heat capacity is $C_\mu = k_B \left(\eta_\mu/ \sinh{\eta_\mu} \right)^2$, with $\eta_\mu=\hbar \omega_\mu/k_B/T/2$. The terms $\hbar \omega_\mu$ and $\mathbf{v}_\mu$ are the phonon frequencies and the group velocities, respectively. In this formulation, the thermal flux is given by $\mathbf{J}(\mathbf{r}) = \left(N_q \mathcal{V}\right)^{-1}\sum_\mu\Delta T_\mu (\mathbf{r}) \mathbf{S}_\mu$, where $\mathcal{V}$ is the volume of the unit cell, and $\mathbf{S}_\mu = C_\mu \mathbf{v}_\mu$. The term $W_{\mu\nu}$ is the energy form of the scattering operator~\cite{romano2020phonon}. While we plan to make available the implementation of Eq.~\ref{bte}, currently only the relaxation-time-approximation (RTA) is supported; within the RTA, the RHS of Eq.~\ref{bte} becomes $C_\mu/\tau_\mu\left[\Delta T_\mu(\mathbf{r}) -  \Delta T(\mathbf{r})\right]$, leading to
\begin{equation}\label{rta}
-\mathbf{v}_\mu \cdot \nabla \Delta T_\mu(\mathbf{r})  = \frac{\Delta T_\mu(\mathbf{r}) - \Delta T(\mathbf{r}) }{\tau_\mu},
\end{equation}
where $\tau_\mu$ is the relaxation time, and $\Delta T(\mathbf{r})$ is the pseudo lattice temperature; this term is evaluated so that energy conservation is satisfied, i.e. $\nabla \cdot \mathbf{J}(\mathbf{r}) = 0$, yielding $\Delta T(\mathbf{r}) = \left[\sum_\nu C_\nu/\tau_\nu \right]^{-1} \sum_\mu C_\mu/\tau_\mu \Delta T_\mu(\mathbf{r})$~\cite{Hua2014TransportConduction}.

The simulation domain is a cuboid of size $L_x$, $L_y$ and $L_z$  with periodic boundary conditions applied throughout by default. A difference of temperature $\Delta T_{\mathrm{ext}}$ is also enforced along a chosen Cartesian axis. At the walls of the pores, energy conservation leads to  $\sum_\nu \Delta T_\nu \mathbf{S}_\mu \cdot\mathbf{\hat{n}} = 0$, which can be split into incoming and outgoing flux, i.e. $\sum_\mu\Delta T_\mu  \mathrm{ReLu}(\mathbf{S}_\mu \cdot\mathbf{\hat{n}}) = -\sum_\mu \Delta T_\mu  \mathrm{ReLu}(-\mathbf{S}_\mu \cdot\mathbf{\hat{n}})$; note that we define $\mathrm{ReLu}(x) = x \Theta(x)$, with $\Theta(x)$ being the Heaviside function. The boundary condition is imposed to phonons leaving the surface; generally, we have $\Delta T_\mu^{-} = \sum_\nu R_{\mu\nu}\Delta T_\nu^+$, where $-$ ($+$) are for outgoing (incoming) phonons, and $R_{\mu\nu}$ is a mode-resolved bouncing matrix. However, currently only a simplified model is available, with which we assume that phonons thermalize to a single temperature~\cite{landon2014deviational}; this choice corresponds to $R_{\mu\nu} = \left[ \sum_k  \mathrm{ReLu}(\mathbf{S}_k  \cdot\mathbf{\hat{n}})\right]^{-1} \mathrm{ReLu}(\mathbf{S}_\mu \cdot\mathbf{\hat{n}})$. Lastly, the effective thermal conductivity, assuming we have imposed the temperature gradient along $x$, reads
\begin{equation}\label{keff}
\kappa^{\mathrm{eff}} = - \frac{L_x}{\Delta T_{\mathrm{ext}}} \sum_\mu  \int_{A}\frac{d\mathbf{S}}{A} \mathbf{J}_\mu\cdot \mathbf{\hat{n}}.
\end{equation}
Equation~\ref{rta}, when solved iteratively, requires inverting as many linear systems as the number of phonons modes, an endeavour that may easily become prohibitive. To ameliorate the computational effort,  we instead implement the recently introduced anisotropic-MFP-BTE (aMFP-BTE), a model based on interpolating the phonon temperatures onto the vectorial MFP space~\cite{romano2021}. In practice, instead of solving Eq.~\ref{rta} $N_pN_q$ times at each iteration, we solve it for a set of vectorial MFPs, $\mathbf{F}_{ml}$, located on a spherical grid. The indices $m$ and $l$ label the MFP $\Lambda_m$ (up to $N_\Lambda$) and phonon directions $\mathbf{\hat{s}}_l$ (up to $N_\Omega$), respectively. As $N_\Lambda N_\Omega << N_p N_q$, while not compromising on accuracy, this scheme results in a much faster runtime with respect to the case with mode-resolved resolution, i.e. no interpolation. 

\section{Implementation}

As derived in~\ref{amfpbte}, the discretization of aMFP-BTE model in real and momentum space yields the following iterative linear system
\begin{eqnarray}\label{iter}
\mathbf{A}_{ml} \mathbf{\Delta T}_{ml}^{(n)} =\sum_{m'l'} \mathbf{S}_{ml}^{m'l'} \mathbf{\Delta T}_{m'l'}^{(n-1)} + \mathbf{P}_{ml},
\end{eqnarray}
where $\mathbf{A}_{ml}$ is the stiffness matrix for a given MFP and direction, $\mathbf{S}_{ml}^{m'l'}$ is associated to the lattice temperature and adiabatic boundary conditions, and $\mathbf{P}_{ml}$ represents the perturbation. Once Eq.~\ref{iter} converges, the effective thermal conductivity is computed as $\kappa_{\mathrm{eff}}=\kappa_{\mathrm{bal}}/2 + \sum_{ml}\mathbf{K}_{ml}^T \cdot \mathbf{\Delta T}_{ml}$, where $\kappa_{\mathrm{bal}}$ is the ballistic thermal conductivity, and $\mathbf{K}_{ml}$ is associated to the thermal flux. 

The first guess to Eq.~\ref{iter} is given by the standard heat conduction equation. While it is straightforward to discretize the diffusive equation in orthogonal grids, it becomes convoluted for unstructured meshes. In \verb!OpenBTE!, we use the approach described in~\cite{murthy2002numerical}, where the non-orthogonal contribution of the flux is treated explicitly, i.e. it depends on previous solutions. The resulting system, analogously to Eq.~\ref{iter}, is an iterative linear system, 
\begin{equation}\label{fourier_iter}
    \mathbf{A} \mathbf{\Delta T}^{(n)}= \mathbf{S} \mathbf{\Delta T}^{(n-1)} + \mathbf{P},
\end{equation}
where $\mathbf{S}$ arises from the non-orthogonality of the mesh. Note that $\mathbf{S}_{ml}^{m'l'}$ in Eq.~\ref{iter} and $\mathbf{S}$ in Eq.~\ref{fourier_iter} are not related. The same applies with the terms sharing the same basename in these two equations. The derivation of Eq.~\ref{fourier_iter} is given in~\ref{fourier}.

Equation~\ref{iter} entails, for each iteration $n$, the inversion of $N_\Lambda N_\Omega$ matrices, whose size is the number of volumes in the computational domain. To enhance the computational efficiency, three distinct strategies have been put in place. First, by noting that the stiffness matrix does not depend on $n$, we are able to compute LU factorizations for the first iteration, and reuse them until convergence. Another major speed up is achieved by vectorizing the assembly of the stiffness matrices, while also exploiting their sparsity. Lastly, for each $n$, Eq.~\ref{iter} consists of a set of independent problems, thus it can be trivially parallelized~\cite{ali2014large}; to this end, we assign a set of phonon directions to each core ($\Omega_c$), while the indices $m$ run serially. The implementations of these three methods rely on \verb!SCIPY!, \verb!NUMPY!~\cite{harris2020array} and \verb!MPI4PY!~\cite{dalcin2011parallel}, respectively. 

When vectorization and parallelization are both used, we may save memory by noting that, as shown in~\ref{amfpbte}, the stiffness matrices have the form $\mathbf{A}_{ml} = \mathbf{I} + \Lambda_m \mathbf{A}_l$; thus vectorization can be done only for $\mathbf{A}_l$, while $\mathbf{A}_{ml}$ is assembled \textit{on the fly} right before LU factorization. Similar arguments hold for $\mathbf{P}_{ml}$ and $\mathbf{S}_{ml}^{m'l'}$. The overall methodology is summerized in Algorithm~\ref{algo}, where, for readability, only $\mathbf{A}_{ml}$ has been expanded upon.
\begin{algorithm}[H]
    \caption{Workflow for thermal conductivity calculations. The term $\Omega_c$ is the set of solid angles assigned to core $c$.}
    \label{algo}
    \begin{algorithmic}[1] 
           \State $\mathbf{\Delta T}_{ml}^{(0)}\leftarrow$ Fourier's law.
           \While{error $>1e^{-3}$} 
                \State $\kappa^{\mathrm{eff},(n+1)} = \kappa_{\mathrm{bal}}/2$
                \For{l in $\{\Omega_c\}$} 
                \For{m = 1:$N_{\Lambda}$}
                 \State $ \mathbf{B} = \sum_{m'l'} \mathbf{S}_{ml}^{m'l'} \mathbf{\Delta T}_{m'l'}^{(n-1)} + \mathbf{P}_{ml}$
                  \State $ \mathbf{\Delta T}_{ml}^{(n)} \leftarrow \left(\mathbf{I} + \Lambda_m \mathbf{A}_l \right)\mathbf{\Delta T}_{ml}^{(n)} = \mathbf{B}  $
                  \State $\kappa^{\mathrm{eff},(n+1)} +=  \mathbf{K}_{ml}^T \cdot \mathbf{\Delta T}_{ml} $
                \EndFor   
                \EndFor   
                
                \State error $\gets |(\kappa^{\mathrm{eff},(n+1)} - \kappa^{\mathrm{eff},(n)})|/\kappa^{\mathrm{eff},(n+1)}$ 
                \State $n \gets n +1$
            \EndWhile\label{euclidendwhile}
    \end{algorithmic}
\end{algorithm}

\section{Workflow}

\begin{figure}
{\includegraphics[width=0.9\textwidth]{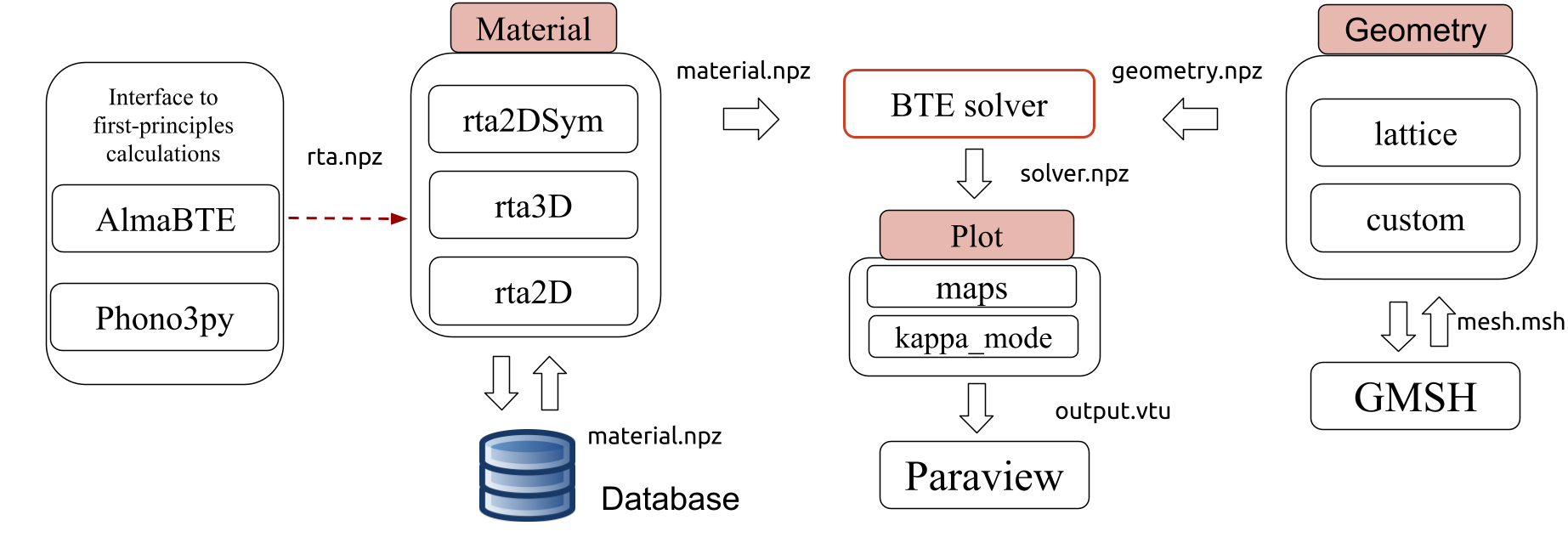}}
\cprotect\caption{The software architecture of \verb!OpenBTE!. The key components are the \verb!Material!, \verb!Geometry!, \verb!Solver! and \verb!Plot! modules. Currently, the code is interfaced to \verb!AlmaBTE! and \verb!Phono3Py!, which computes first-principles data. Mesh generation is delegated to \verb!GMSH!, and output maps can be visualized with \verb!Paraview!.\label{fig11}}
\end{figure}

The easiest way to run \verb!OpenBTE! is to use it as a module in \verb!Python!. A sketch for the workflow is illustrated in Fig.~\ref{fig11}. The core module is \verb!Solver!, which runs both the heat conduction equation as a first guess to Eq.~\ref{iter} and all subsequent iterations, outlined in Algorithm~\ref{algo}. Before invoking it, though, it is necessary to define the material's internal boundaries using the \verb!Geometry! module, and bulk-related data via the \verb!Material! module. The latter is the interface to external packages. Once the simulation is finalized, results can be plotted using the \verb!Plot! module. Note that \verb!OpenBTE! follows a pure functional approach, where data is simply stored as dictionaries. Each module is expanded upon in the next sections.

\section{Material}

The \verb!Material! module converts mode-resolved bulk data, $C_\mu$, $\tau_\mu$ and $\mathbf{v}_\mu$, into the spherically-resolved quantities $\mathbf{G}_{ml}$ and $C_{ml}$.  The input files, \verb!rta.npz!, are created by postprocessing data obtained by external packages. Currently, \verb!OpenBTE! provides interfaces to \verb!AlmbaBTE!~\cite{Carrete2017AlmaBTEMaterials} and \verb!Phono3Py!~\cite{phono3py}. A small set of precomputed \verb!rta.npz! is also provided. The file created by \verb!Material! is stored into \verb!material.npz! and ready to be used by \verb!Solver!. Depending on the geometry, this module has two models:
\begin{itemize}
    \item \verb!rta3D!: this model must be used for geometries with $l_z > 0$. In this case, the angular interpolation of $\mathbf{F}_\mu$ is performed in both polar and azimuthal angles.
     \item \verb!rta2DSym!: when $l_z$ is not specified, the actual simulation domain is 2D, and this model must be used. Note that when the base material is 3D, using this option is equivalent to considering an infinite thickness. In fact, in this case, the mode-resolved $\mathbf{F}_\mu$ are mapped into $\mathbf{F}_{tk}=\Lambda_t \cos(\phi_k)$, where $\Lambda_t$ implicitly includes the MFP and the azimuthal angle, and $\phi_k$ is the polar angle on the \textit{x-y} plane~\cite{romano2021}. 
\end{itemize}
The model as well as the discretization grid can be chosen with

\begin{minted}[breaklines]{python}
from openbte import Material

Material(filename='rta_Si_300',model='rta2DSym',n_phi=48)
\end{minted}
where, by default, the material file is taken from the database, whose entries are updated online. In this case, we choose to discretize the polar angle into 48 slices.

\section{Geometry module}
The \verb!Geometry! module allows the user to build an arbitrarily patterned nanomaterial. It interfaces with \verb!GMSH!~\cite{geuzaine2009gmsh} to create an unstructured Delaney mesh, and creates the file \verb!geometry.npz!. While we plan to handle user-provided meshes, \verb!OpenBTE! currently supports 2D membranes with finite and infinite thickness. The size of the unit cell is specified by \verb!lx! and \verb!ly!, while the thickness is defined by \verb!lz!. The direction of the applied gradient is indicated with \verb!direction! (default is \verb!x!). Periodic boundary conditions are applied along all directions; however, adiabatic surfaces can be imposed along specific directions with \verb!Periodic=[True,False,True]!, where each component declares whether a surface is periodic. Currently, there are two geometry models, as expanded upon in the next sections.

\subsection{Lattice model}

This model aids the creation of 2D periodic porous materials with pores having either predefined shapes or custom ones. In analogy to crystals, a porous material is defined as a set of unit vectors, which in our case are $\mathbf{a}_1 = l_x \mathbf{\hat{x}}$ and $\mathbf{a}_2 = l_y \mathbf{\hat{y}}$, and a base. The latter is a list of \textit{x-y} pairs, each of them being one pore. Here is an example with a single pore in the origin
    
\begin{minted}[breaklines]{python}
from openbte import Geometry

Geometry(model='lattice',base=[[0,0]],shape='circle')
\end{minted}

In the example above, we chose a simple shape, a circle; however, it is possible also to supply a user-defined shape using \verb!shape=custom!. In such a case, the keyword \verb!shape_functions! must be provided, optionally along with \verb!shape_options!. 
For example, as shown in Fig.~\ref{fig:shape}a-b and c, we can define a structure having two pores, each of them described by the same custom shape.


\subsection{Custom model}
    
\begin{figure}
\begin{center}
{\includegraphics[width=1\textwidth]{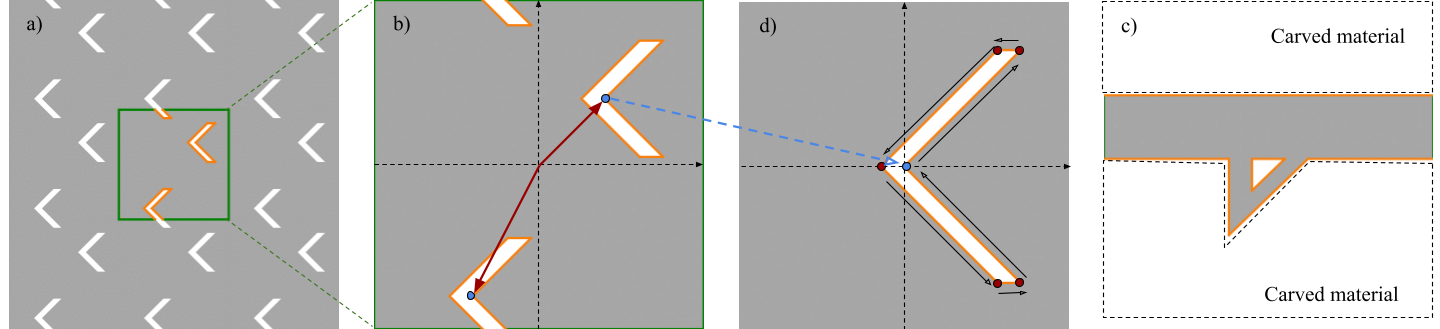}}
\cprotect\caption{a) An example of structure using the keyword-value pair \verb!shape=custom! for the \verb!lattice! model of the \verb!Geometry! module. b) the unit cell comprising two pores at designed positions. Note that when a pore crosses a boundary it gets repeated. c) singe-pore view of material. The coordinates of the shape are defined via the \verb!shape_function! option. d) An example of \verb!custom! shape obtained by carving out the base material with user-defined polygons.  \label{fig:shape}  }
\end{center}
\end{figure}

    Using the option \verb!model=custom!, it is possible to define an arbitrary patterning by providing a list of polygons, with the periodicity being automatically guaranteed. This option significantly widens up the number of possible geometries at the expense of more input by the user. In spirit of \textit{subtractive manufacturing}, each polygon can be seen as a region where the material is carved out. An example of custom model is depicted in Fig.\ref{fig:shape}-d.

\section{Plot Module}

Simulations results can be plotted via the \verb!Plot! module. Currently, the following features are implemented:
\begin{itemize}
    \item \verb!vtu!: the temperature and flux maps are stored in a \verb!vtu! file and ready to be opened by \verb!Paraview!~\cite{ahrens2005paraview}.
    \item \verb!maps!: the maps are shown as web-ready, interactive structures.  
    \item \verb!kappa_mode!: this model computes the mode-resolved thermal conductivity from Eq~\ref{mode-resolved}, and stores in a file.
    \item \verb!line!: this option, which currently is implemented only for 2D systems, allows for line plots, using linear interpolation.
    
\end{itemize}

\section{Thermal transport in porous Si}

\begin{figure*}[t!]
\subfloat[\label{fig1a}]
{\includegraphics[width=0.5\textwidth]{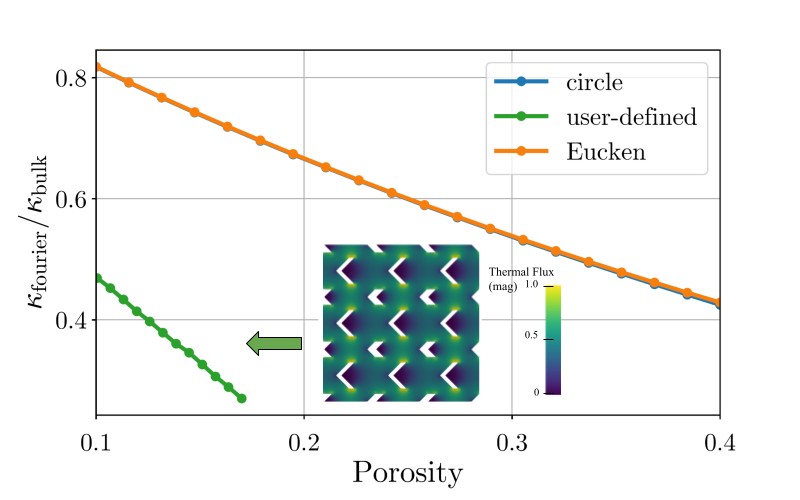}}
\hfill
\subfloat[\label{fig1b}]
{\includegraphics[width=0.5\textwidth]{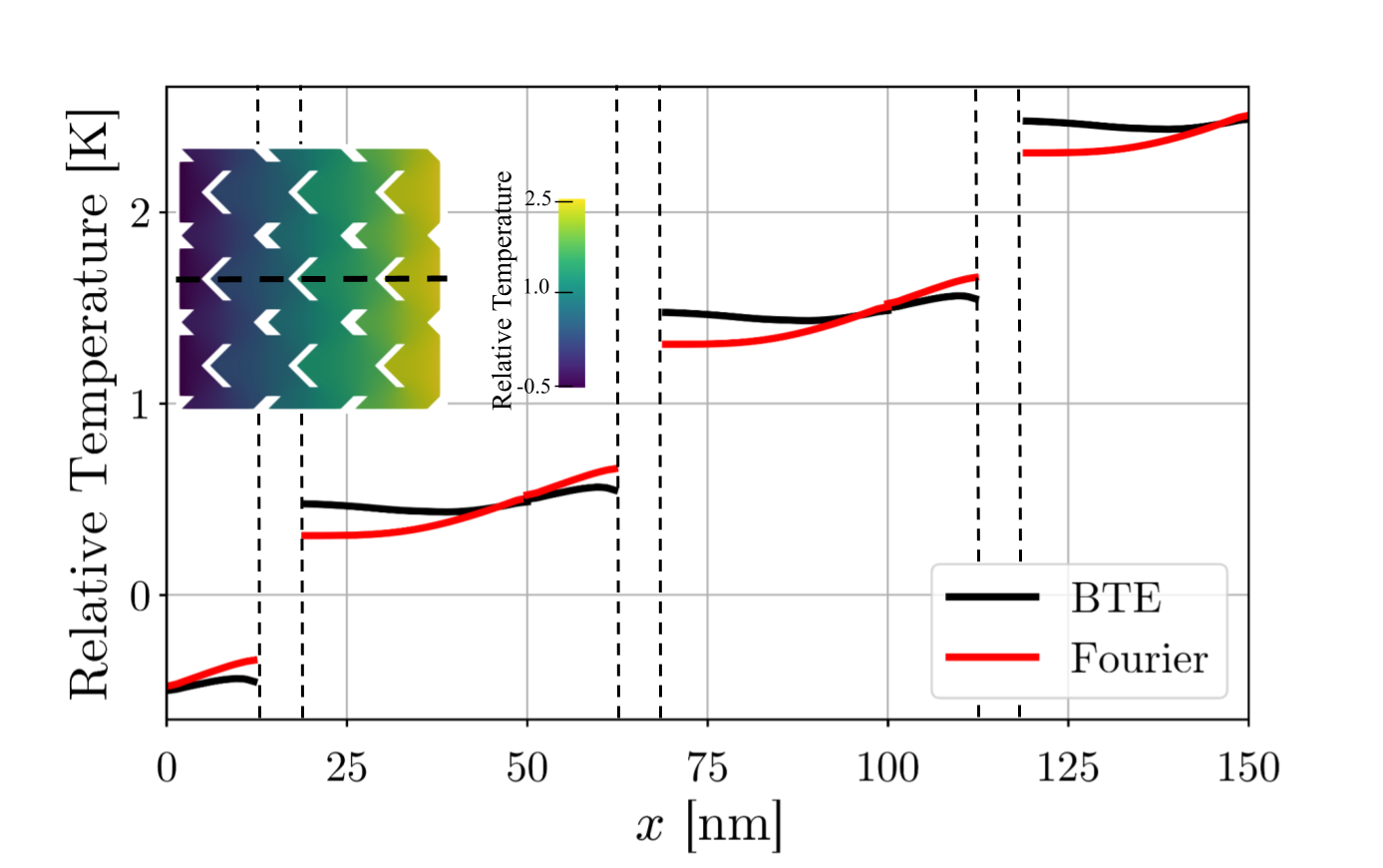}}
\quad
\subfloat[\label{fig1c}]
{\includegraphics[width=0.5\textwidth]{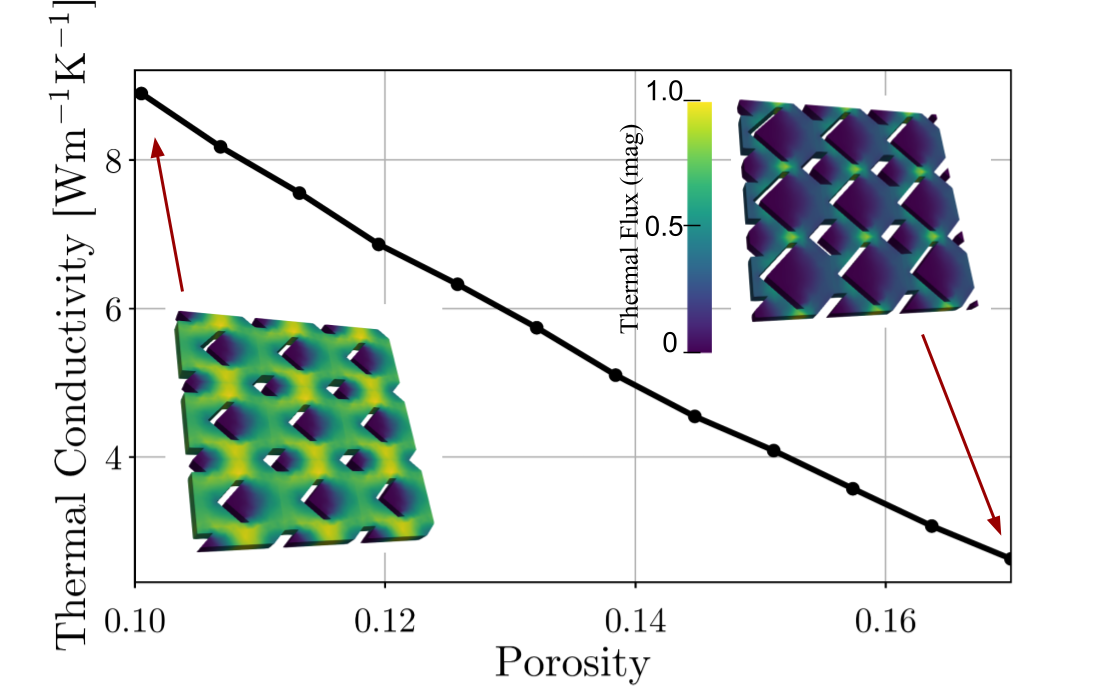}}
\hfill
\subfloat[\label{fig1d}]
{\includegraphics[width=0.5\textwidth]{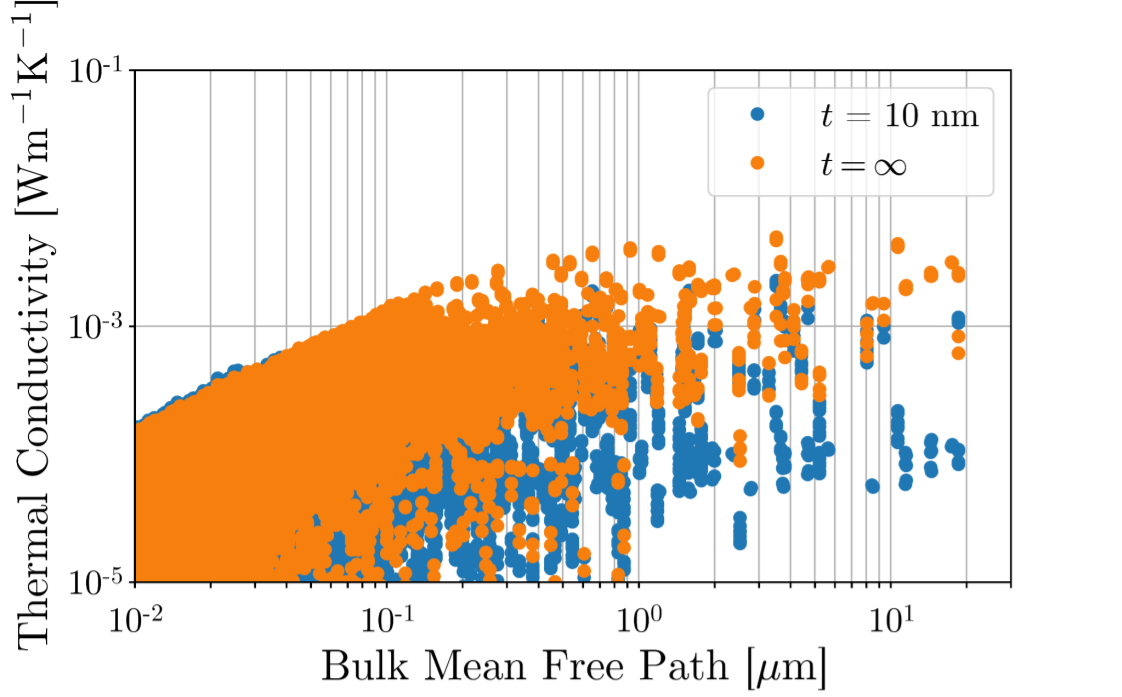}}
\caption[]{a) Thermal transport reduction when using Fourier's law, for regular circular pores and for a user-define shape; the Eucken-Maxwell model is also plotted. In the inset, the magnitude of the thermal map for the user-defined structure. The porosity is 0.1 and $L$ = 50 nm. b) a cut of the temperature maps, computed from both the BTE and Fourier's law, along the $x$-axis. c) The effective thermal conductivity, $\kappa^{\mathrm{eff}}$, of the user-defined structure with thickness $t$ = 10 nm, and for different porosities. In the inset, the case with porosities = 0.1 and 0.17 are shown. d) The mode-resolved thermal conductivity, $\kappa^{\mathrm{eff}}_\mu$, versus the MFP, $|F_{\mu,x}|$, for the cases $t$ = 10 nm and $t = \infty$.}
\end{figure*}

 In some cases, it is convenient to use Fourier's law to estimate macroscopic geometric effects. In this first part, we thus aim to assess the reliability of the diffusive solver, implemented in \verb!OpenBTE!. To this end, we consider a membrane with circular pores and infinite thickness, for which $\kappa_\mathrm{fourier}/\kappa_{\mathrm{bulk}} = r$ can be evaluated analytically. In fact, in this case, the thermal transport suppression is given by the Eucken-Maxwell model $r = (1-\phi)/(1+\phi) $~\cite{hasselman1987effective}, $\phi$ being the porosity. In Fig.~\ref{fig1a}, we compare this model with the results from \verb!OpenBTE! for different porosities, obtaining excellent agreement ($<1\%$). We note that by using the option \verb!only_fourier=True! it is possible to run only the Fourier's model. Next, we compute $r$ for a more complicated structure, comprising two concatenated sub-lattices of pores with different user-provided shapes; this structure is generated using the \verb!shape=custom! option from the \verb!lattice! model, a two-pore base, and different options for each pore. As shown in Fig.~\ref{fig1a}, we obtain a much larger heat transport suppression due to the longest path heat must travel through the material. 
 
 Let us now compute $\kappa^{\mathrm{eff}}$ using the BTE, encoded in Eq.~\ref{iter}. For the chosen porosity of $0.1$ and $L$ = 50nm, we obtain ~17.4 Wm$^{-1}$K$^{-1}$, well below the corresponding macroscopic value~99.4 Wm$^{-1}$K$^{-1}$. In Fig.~\ref{fig1b}, we plot a cut of the temperature maps, both for the BTE and Fourier cases, along the $x$-axis. As expected, they have different trends, the BTE one being flatter due to ballistic transport. When a finite thickness is added, phonon undergo stronger suppression; in fact, for thickness $t$=10 nm and porosity 0.1, we obtain $\kappa^{\mathrm{eff}} = 8.9$ W m$^{-1}$K$^{-1}$, roughly a half of the value with $t = \infty$. In Fig.~\ref{fig1c}, $\kappa^{\mathrm{eff}}$ for different porosities is plotted. From the insets, we note that, as the porosity becomes larger, the heat flux peaks around the phonon bottleneck, e.g. the areas between the pores. Lastly, the mode-resolved thermal conductivities for $t$ = 10 nm and $\infty$ are plotted in Fig.~\ref{fig1d}; analogously to Ref.~\cite{romano2021}, we can appreciate the stronger suppression for large MFPs for the case with finite thickness. 

\section{Conclusion}

We have presented \verb!OpenBTE!, a software for computing deterministically space-dependent heat transport in 2D and 3D structures. The efficient and modern implementation of the underlying algorithms enable accurate simulations both in the cloud and on common laptops, thus democratizing this type of simulations. Future work includes going beyond the RTA of the BTE and adding transient dynamics. GPU support and differentiability are also on the roadmap. Furthermore, we plan to develop a hybrid Fourier/BTE solver, which has the potential to dramatically reduce runtimes for the cases with 3D structures. Filling an important gap in the current offer of publicly available phonon transport solvers, \verb!OpenBTE! sets out to guide experiments on nanoscale heat transport and facilitate high-throughput prediction of thermal properties of nanostructures. 

\section{Acknowledgments}

Research  was  partially  supported  by  the  Solid-State SolarThermal  Energy  Conversion  Center  (S3TEC), an Energy Frontier Research Center funded by the U.S. Department of Energy (DOE), Office of Science, Basic Energy Sciences (BES), under Award No.  DESC0001. The author thanks all the early users, who helped improving \verb!OpenBTE!.

\begin{figure}
\begin{center}
{\includegraphics[width=0.6\textwidth]{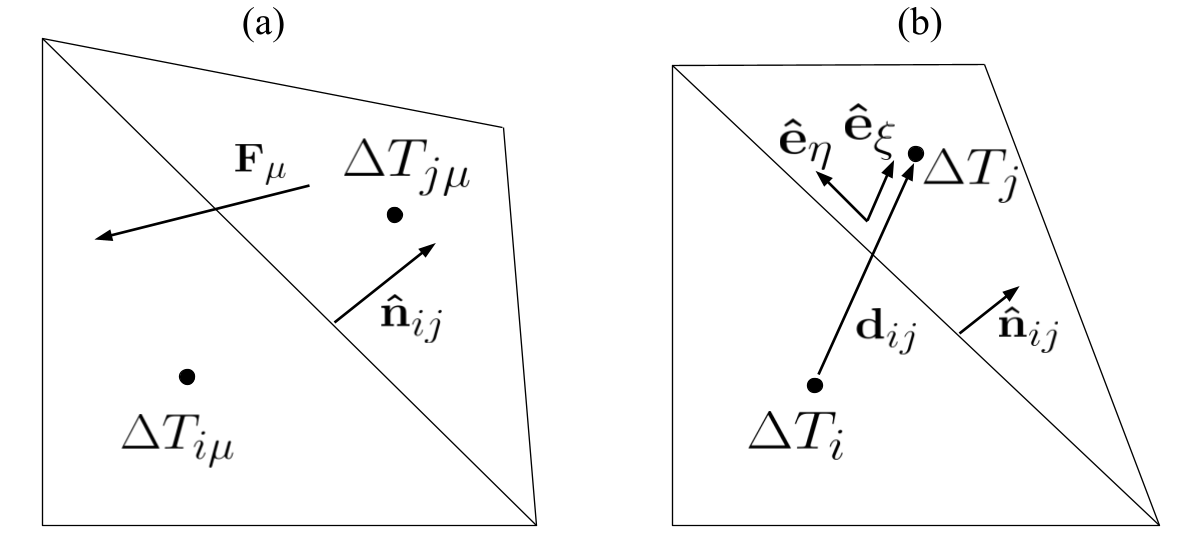}}
\caption{a) A scheme for two adjacent elements. In this case, the vectorial MFP $\mathbf{F}_\mu$ points inward the element $i$, thus it contributes to the off-diagonal part of the stiffness matrix of the BTE model. b) Local reference system, $\mathbf{\hat{e}}_\xi-\mathbf{\hat{e}}_\eta$, used to derive the non-orthogonal contribution to the heat flux for the heat-conduction equation. \label{fig3}}
\end{center}
\end{figure}

\appendix
\section{The aMFP-BTE}
\label{amfpbte}

\verb!OpenBTE! employs the upwind finite-volume discretization of Eq.~\ref{rta}, where the flux outgoing from volume $i$ contributes to the diagonal of the stiffness matrix at row $i$~\cite{murthy2002numerical}. A sketch depicting incoming and outgoing flux is illustrated in Fig.~\ref{fig3}a. The  discretized mode-resolved BTE reads~\cite{romano2021}
\begin{equation}\label{sylvester}
\sum_{j\mu'} \left[ \delta_{cc'}\delta_{\mu\mu'} + A_{ij\mu'} + B_{\mu\mu'j}\delta_{ij} + D_{\mu'}\delta_{ij} \right] \Delta T_{j\mu'} =  P_{i\mu},
\end{equation}
where
\begin{eqnarray}
A_{ij\mu} &=& \sum_{k} \mathrm{ReLu}(\mathbf{F}_\mu\cdot \mathbf{R}_{ik} )  \delta_{ij} -\mathrm{ReLu}(-\mathbf{F}_\mu\cdot \mathbf{R}_{ik} ) \delta_{jk},\\
B_{\mu\mu'j} &=& -\sum_{s} \mathrm{ReLu}(-\mathbf{F}_\mu\cdot \mathbf{\hat{n}}_{s} ) g_{si} \frac{\mathrm{ReLu}(\mathbf{S}_\nu\cdot\mathbf{\hat{n}}_s) }{\sum_k\mathrm{ReLu}(\mathbf{S}_k\cdot\mathbf{\hat{n}}_s)},
\end{eqnarray}
arise flux discretization and adiabatic boundary conditions, respectively. The term $D_\mu = C_\mu/\tau_\mu \left[\sum_k C_k/\tau_k \right]^{-1} $ contributes to the lattice pseudo-temperature; lastly, the perturbation is 
\begin{equation}
P_{i\mu} = -\sum_{j}  \mathrm{ReLu}(-\mathbf{F}_\mu\cdot \mathbf{R}_{ij}^P ) \Delta T_{\mathrm{ext}}.
\end{equation}
The terms $\mathbf{R}_{ij}$,  $\mathbf{R}_{ij}^P$ and $g_{sc}$ define the mesh, and are given by 
\begin{eqnarray}
\mathbf{R}_{ij} &=  &\begin{cases}
   \frac{1}{V_c}A_{ij}\mathbf{\hat{n}}_{ij},& \text{if } i \text{ and } j \text{ are neighbors}\\
    0,              & \text{otherwise},
\end{cases}\\
\mathbf{R}_{ij}^P& =  &\begin{cases}
   \frac{1}{V_i}A_{ij}\mathbf{\hat{n}}_{ij},& \text{if } i \text{ and } j \text{ are periodic to each other}\\
    0,              & \text{otherwise}.
\end{cases}\\
g_{si} &=  &\begin{cases}
   \frac{1}{V_{i}}A_s ,& \text{if } s \text{ is a side of } i\\
    0,              & \text{otherwise},
\end{cases}
\end{eqnarray}
where $A_{ij}$ is the area of the surface between the volumes $i$ and $j$, and $\mathbf{\hat{n}}_{ij}$ is its normal, pointing toward the volume $j$. The term $V_i$ is the volume of the element $i$. 
We solve~\ref{sylvester} iteratively, 
\begin{equation}\label{iter3}
\sum_{j} \left(\delta_{ij} + A_{ij\mu}\right) \Delta T_{j\mu}^{(n)} = P_{i\mu} + \sum_{\mu'}\left( B_{\mu\mu'i} + D_{\mu'} \right) \Delta T_{i\mu'}^{(n-1)}.
\end{equation}
Equation~\ref{iter3} requires solving as many linear systems as the number of phonons modes. To overcome this issue, \verb!OpenBTE! implements the aMFP-BTE~\cite{romano2021}, where the phonon temperatures are interpolated over the vectorial MFPs, $\mathbf{F}_\mu = \sum_{ml} c_{ml}^{\mu} \mathbf{F}_{ml}$; the chosen $\mathbf{F}_{ml} =\Lambda_m \mathbf{\hat{s}}_l$ are located uniformnly on a spherical grid. The labels $m$ and $l$ refer to MFP and solid angle, respectively. Analogously, the temperatures are $\mathbf{\Delta T}_\mu = \sum_{ml} c_{ml}^{\mu} \mathbf{\Delta T}_{ml}$. In the aMFP-BTE formulation, Eq.~\ref{iter3} becomes
\begin{equation}\label{iter2}
\sum_{j} \left(\delta_{ij} + \Lambda_m A_{ijl}\right) \Delta T_{jml}^{(n)} = \Lambda_m P_{il} + \sum_{m'l'}\left( \Lambda_m B_{lm'l'i} + D_{m'l'} \right) \Delta T_{im'l'}^{(n-1)},
\end{equation}
where
\begin{eqnarray}
A_{ij'l} &=& \sum_{k} \mathrm{ReLu}(\mathbf{\hat{s}}_l\cdot \mathbf{R}_{ik} )  \delta_{ij} -\mathrm{ReLu}(-\mathbf{\hat{s}}_l\cdot \mathbf{R}_{ik} ) \delta_{jk},\\
B_{lm'l'j} &=& -\sum_{s} \mathrm{ReLu}(-\mathbf{\hat{s}}_l\cdot \mathbf{\hat{n}}_{s} ) g_{si} \frac{\mathrm{ReLu}(\mathbf{S}_{m'l'}\cdot\mathbf{\hat{n}}_s) }{\sum_{m''l''}\mathrm{ReLu}(\mathbf{S}_{m''l''}\cdot\mathbf{\hat{n}}_s)}.
\end{eqnarray}
The term $P_{il}$ is the perturbation, given by
\begin{equation}
P_{il}= -\sum_{j}  \mathrm{ReLu}(-\mathbf{\hat{s}}_l\cdot \mathbf{R}_{ij}^P ) \Delta T_{\mathrm{ext}}.
\end{equation}
The quantities $D_{ml}=\sum_{\mu} c_{ml}^{\mu} D_\mu$ and $\mathbf{S}_{ml} = \sum_\mu c_{ml}^{\mu} \mathbf{S}_\mu$ contributes to the lattice temperature and thermal flux, respectively. Finally, after defining $S_{iml}^{m'l'} = \Lambda_m B_{lm'l'i} + D_{m'l'}$ in Eq.~\ref{iter2}, we obtain Eq.~\ref{iter}. 
To compute the effective thermal conductivity from Eq.~\ref{keff}, we need to integrate over the hot contact and the momentum space. According to the upwind scheme, the volume from which we pick the temperature depends on whether the flux is outgoing from or incoming to such volume; accordingly, Eq.~\ref{keff} translates into
\begin{eqnarray}
    \kappa_{\mathrm{eff}} = &-&\frac{L}{A \Delta T_{\mathrm{ext}}}\sum_{ijml}  \big[ \Delta T_{cml}  \mathrm{ReLu}(\mathbf{S}_{ml}\cdot\mathbf{R}^K_{ij}) - \nonumber\\ &-&\left(\Delta T_{iml}  + \Delta T_\mathrm{ext}\right) \mathrm{ReLu}(-\mathbf{S}_{ml}\cdot \mathbf{R}^K_{ij}) \big] =  \nonumber \\
    &=& \frac{\kappa_{\mathrm{bal}}}{2} + \sum_{ml}\mathbf{K}_{ml}^T\mathbf{\Delta T}_{ml} 
\end{eqnarray}
where $\kappa_{\mathrm{bal}} = L \sum_{ml} \mathbf{G}_{ml}\cdot\mathbf{\hat{n}_{\mathrm{ext}}}$ is the ballistic thermal conductivity, $\mathbf{\hat{n}}_{\mathrm{ext}}$ being the direction of the applied temperature gradient; the matrices $\mathbf{K}_{ml}$ are
\begin{equation}
K_{iml} = \frac{L}{A \Delta T_{\mathrm{ext}}}\sum_{j} \mathbf{S}_{ml}\cdot\mathbf{R}^K_{ij},
\end{equation}
where we use $\mathrm{ReLu}(x) - \mathrm{ReLu}(-x) = x$, and
\begin{eqnarray}
\mathbf{R}_{ij}^K &=  &\begin{cases}
   A_{ij}\mathbf{\hat{n}}_{ij},& \text{if } i \text{ and } j \text{ share a hot contact's side}\\
    0,              & \text{otherwise}.
\end{cases}
\end{eqnarray}
Lastly, it is also possible to compute the ``mode-resolved'' thermal conductivity, given by
\begin{equation}\label{mode-resolved}
    \kappa^{\mathrm{eff}}_\mu = \frac{\kappa_{\mathrm{bal}}}{2} + \sum_{ml}\mathbf{K}_{\mu}^T c_\mu^{ml}\mathbf{\Delta T}_{ml},
\end{equation}
where $c_\mu^{ml}$ maps $\mathbf{F}_{ml}$ to $\mathbf{F}_{\mu}$, and
\begin{equation}
K_{i\mu} = \frac{L}{A \Delta T_{\mathrm{ext}}}\sum_{j} \mathbf{S}_\mu\cdot\mathbf{R}^K_{ij}.
\end{equation}

\section{Heat conduction equation}
\label{fourier}

The first guess to Eq.~\ref{iter} is given by the standard diffusive equation $\nabla \kappa \nabla \Delta T (\mathbf{r})= 0$, with $\kappa$ being the bulk thermal conductivity tensor. In a finite-volume fashion, we integrate over the control volume $\Omega_c$
\begin{equation}\label{balance}
   \int_{\partial \Omega_c} \sum_j \kappa_{ij}\frac{\partial \Delta T(\mathbf{r})}{\partial x_j}  \hat{n}_i = 0,
\end{equation}
where we used Gauss' law. Eq.~\ref{balance} is then transformed into a sum of integrals over individual faces of the volume, i.e. 
\begin{equation}\label{balance2}
   \sum_j \mathbf{g}_{ij} \cdot \mathbf{S}_{ij} = 0
\end{equation}
where $S_{ij} = \sum_\beta \kappa_{\alpha\beta} R_{ij,\alpha}$ and $\mathbf{g}_{ij}$ is the gradient evaluated at the face between the volume $i$ and $j$ (which hereafter will be referred to as $s$). For orthogonal grids, flux discretization simply leads to $\nabla T = \left(\Delta T_j-\Delta T_i\right)\mathbf{d_{ij}}|\mathbf{d_{ij}}|^{-2}$, where $\mathbf{d}_{ij}$ is the distance between the centroids of the two adjacent volumes. However, for unstructured mesh, a non-orthogonal contribution must be added. In \verb!OpenBTE!, we follow the approach described by Murthy~\cite{murthy2002numerical}, described as follows. With no loss of generality, we consider a 2D case. As shown in Fig.~\ref{fig3}b, we introduce a reference system based on the vectors $\mathbf{e}_\eta$ and $\mathbf{e}_\xi$. The former is aligned with $\mathbf{d}_{ij}$ and latter is parallel to the face. In this new system, the gradient transforms as
\begin{equation}
   \frac{\partial \Delta T}{\partial \mathbf{x}} = \mathcal{J}(\mathbf{x},\mathbf{e}) \frac{\partial \Delta T}{\partial \mathbf{e}},
\end{equation}
with $\mathcal{J}$ being the Jacobian. Here we will not elaborate on its structure but will only provide the final formula for the temperature gradient~\cite{murthy2002numerical} 
\begin{equation}\label{grad}
 \mathbf{g}_{ij}= \frac{\mathbf{
 \hat{n}}_{ij}}{\mathbf{\hat{n}}_{ij}\cdot \mathbf{\hat{e}}_\xi} \left[  \frac{\partial \Delta T}{\partial e_\xi}\bigg|_{s} - \mathbf{\hat{e}}_\xi \cdot \mathbf{\hat{e}}_\eta \frac{\partial \Delta T}{\partial e_\eta}\bigg|_{s}\right].
\end{equation}
The gradient along $\mathbf{e}_\xi$ at the face $s$ is simply 
\begin{equation}\label{ortho}
    \frac{\partial \Delta T}{\partial e_\xi}\bigg|_{s} = \frac{\Delta T_j - \Delta T_i - R_{ij}^P}{|\mathbf{d_{ij}}|},
\end{equation}
where
\begin{eqnarray}
R_{ij}^P &=  &\begin{cases}
   -\Delta T_{\mathrm{ext}},& \text{if } i \in \mathrm{hot} \text{ and } j \in \mathrm{cold}  \text{ and they share a side}  \\
   \Delta T_{\mathrm{ext}},& \text{if } i \in \mathrm{cold} \text{ and } j \in \mathrm{hot} \text{ and they share a side} \\
   0 & \text{otherwise}
\end{cases}
\end{eqnarray}
Using Eq.~\ref{ortho}, we rewrite Eq.~\ref{grad} as
\begin{equation}\label{grad5}
 \mathbf{g}_{ij} = \frac{\mathbf{\hat{n}}_{ij}}{\mathbf{\hat{n}}_{ij}\cdot \mathbf{d}_{ij}} \left[  \Delta T_j - \Delta T_i - R_{ij}^P\right] + \mathbf{F}_{ij},
\end{equation}
where $\mathbf{F}_{ij}$ is called ``secondary flux.'' In contrast, Eq.~\ref{ortho} is termed ``primary flux.'' The secondary flux is treated explicitly, i.e. using precomputed temperatures. To avoid unambiguities in the 3D case, we compute $\mathbf{F}$ as the difference between the total flux and the primary flux, i.e.
\begin{equation}\label{grad3}
\mathbf{F}_{ij}=\mathbf{g}_{ij}- \frac{\mathbf{\hat{n}}_{ij}}{\mathbf{\hat{n}}_{ij}\cdot \mathbf{d}_{ij}} \left( \mathbf{g}_{ij} \cdot \mathbf{d}_{ij}\right).
\end{equation}
Combining Eqs.~\ref{grad3}-\ref{grad5}, and explicitly indicating the iteration index, we have
\begin{equation}\label{grad2}
\mathbf{g}_{ij}^{(n)}= \mathbf{g}_{ij}^{(n-1)} + \frac{\mathbf{\hat{n}}_{ij}}{\mathbf{\hat{n}}_{ij}\cdot \mathbf{d}_{ij}} \left[\left(\Delta T_j^{(n)}-\Delta T_i^{(n)} \right)-\mathbf{g}_{ij}^{(n-1)}\cdot \mathbf{d}_{ij}\right].
\end{equation}
In light of these results, Eq.~\ref{balance2} translates into the following linear system 
\begin{equation}\label{preiter}
    \mathbf{A} \mathbf{\Delta T}= \mathbf{B}+ \mathbf{P},
\end{equation}
with stiffness matrix and right hand side 
\begin{eqnarray}\label{final}
     A_{ij} &=& \mathbf{S}_{ij}\cdot  \frac{\mathbf{n}_{ij}}{\mathbf{\hat{n}}_{ij}\cdot \mathbf{d}_{ij}}    -\delta_{ij}\sum_{k} \mathbf{S}_{ik}\cdot    \frac{\mathbf{n}_{ik}}{\mathbf{\hat{n}}_{ij}\cdot \mathbf{d}_{ik}}, \nonumber \\
     B_i &=& \sum_{j} \mathbf{g}_{ij} \cdot \left( \frac{\mathbf{d}_{ij}(\mathbf{\hat{n}}_{ij}\cdot \mathbf{S}_{ij})}{\mathbf{\hat{n}}_{ij}\cdot \mathbf{d}_{ij}} -\mathbf{S}_{ij}\right).
\end{eqnarray}
The perturbation term is given by
\begin{equation}\label{pert2}
    P_i = \sum_{j}\mathbf{S}_{ij}^P\cdot \frac{\mathbf{n}_{ij}}{\mathbf{\hat{n}}_{ij}\cdot \mathbf{d}_{ij}},
\end{equation}
where
$\mathbf{S}_{ij}^P = \mathbf{S}_{ij}R_{ij}^P$. The gradients $\mathbf{g}_{ij}$, appearing in Eq.~\ref{final}, depend on precomputed temperatures. Their calculation if performed by interpolating the gradients from the elements $i$ and $j$, i.e.
\begin{equation}
    \mathbf{g}_{ij}=\omega_{ij} \mathbf{g}_i+ \left(1-\omega_{ij} \right)\mathbf{g}_j,
\end{equation}
where
\begin{equation}
     \mathbf{g}_i = \frac{\partial \Delta T}{\partial \mathbf{x}}\bigg|_{i},
\end{equation}
and $\omega_{ij}$ is the distance between the centroid of the element $i$ and the intersection between $\mathbf{d}_{ij}$ and the face $s$. The gradient at the element's centroid is evaluated assuming local linear variation of $\Delta T$ around the element $i$, resulting in as many linear equations as the neighbours of $i$,
\begin{equation}
\Delta T_i + \mathbf{g}_i\cdot \mathbf{d}_{ij}  = \Delta T_j + R_{ij}^P,
\end{equation}
which, in matrix form, become
\begin{equation}\label{lq}
 \mathbf{M}_i \mathbf{g}_i  = \mathbf{Q}_i.
\end{equation}
Equation~\ref{lq} is a linear system to be solved for each element. Since $\mathbf{M}_i$ has more rows (number of neighbors) that columns (dimensions), we solve it using the least square method, i.e.
\begin{equation}\label{mm}
   \mathbf{g}_i  =  \left(\mathbf{M}_i^T\mathbf{M}_i\right)^{-1} \mathbf{M}_i^T \mathbf{Q}_i.
\end{equation}
Combining Eqs.~\ref{mm}-\ref{final} and-\ref{preiter}, we may formalize the heat conduction equation in unstructured grids as the following iterative linear system
\begin{equation}\label{final2}
    \mathbf{A} \mathbf{\Delta T}^{(n)}= \mathbf{S} \mathbf{\Delta T}^{(n-1)} + \mathbf{P},
\end{equation}
as reported in the main text. Lastly, after Eq.~\ref{final2} is solved, $\kappa_{\mathrm{fourier}}$ is computed by 
\begin{equation}
    \kappa_{\mathrm{fourier}} = -\frac{L}{\Delta T_{\mathrm{ext}}A}\sum_{s\in\mathrm{hot}}\mathbf{g}_{ij}\cdot \mathbf{S}_{ij}.
\end{equation}


\bibliographystyle{elsarticle-num}

\bibliography{biblio.bib}







\end{document}